\documentclass[structabstract]{aa}
\usepackage{graphicx}
\usepackage{txfonts}
\usepackage{natbib}
\bibpunct{(}{)}{;}{a}{}{,}
\usepackage[bookmarks=false, colorlinks=true, citecolor=blue, linkcolor=blue]{hyperref}
\usepackage{lscape}
\usepackage{longtable}
\usepackage{amssymb}
\usepackage[Symbol]{upgreek}
\usepackage[T1]{fontenc}
\def\kms{km\,s$^{-1}$}
\def\Jykms{Jy\,km\,s$^{-1}$}

\begin{document}

\title{Giant burst of methanol maser in S255IR-NIRS3}
  \titlerunning{Giant burst of methanol maser}
  \authorrunning{M. Szymczak et al.}

\author{M. Szymczak\inst{\ref{inst1}} \and M. Olech\inst{\ref{inst1}} \and P. Wolak\inst{\ref{inst1}} \and E. G\'erard\inst{\ref{inst2}} \and A. Bartkiewicz\inst{\ref{inst1}}}
\institute{Centre for Astronomy, Faculty of Physics, Astronomy and Informatics, Nicolaus Copernicus University, \\ Grudziadzka 5, PL-87100 Torun, Poland\label{inst1}
\and
GEPI, UMR 8111, CNRS and Observatoire de Paris, 5 Place J. Janssen, F-92195 Meudon Cedex, France\label{inst2}
}
\date{Received 17 May 2018 / Accepted 12 July 2018}

\abstract
{High-mass young stellar objects (HMYSOs) can undergo accretion episodes that strongly affect the star evolution,
the dynamics of the disk, and its chemical evolution. Recently reported extraordinary bursts in the methanol maser emission
may be the observational signature of accretion events in deeply embedded HMYSOs.}
{We analyze the light curve of 6.7\,GHz methanol masers in S255IR-NIRS3 during the 2015-2016 burst.}
{8.5-year monitoring data with an average sampling interval of 5 days were obtained with the Torun 32\,m radio telescope.
Archival data were added, extending the time series to $\sim$27\,years.}
{The maser emission showed moderate (25-30\%) variability on timescales of months to years over $\sim$23\,years since its discovery.
The main burst was preceded by a one-year increase of the total flux density by a factor of 2.5, then it
grew by a factor of 10 over $\sim$0.4\,years and declined by a factor of 8 during the consecutive 2.4 years. The peak maser luminosity
was a factor of 24.5 higher than the pre-burst quiescent value. The light curves of individual features showed considerable diversity
but indicated a general trend of suppression of the maser emission at blueshifted ($<$4.7\,\kms)
velocities when the redshifted emission rapidly grew and new emission features appeared at velocities $>$5.8\,\kms. This new emission
provided a contribution of about 80\% to the maser luminosity around the peak of the burst. The duration of the burst at the extreme redshifted
velocities of 7.1 to 8.7\,\kms\ was from 0.9 to 1.9\,years, and its lower limit for the other features was $\sim$3.9\,years.}
{The onset of the maser burst exactly coincides with that of the infrared burst estimated from the motion of the light echo. This
strongly supports the radiative pumping scheme of the maser transition. The growth of the maser luminosity is the result of an increasing
volume of gas where the maser inversion is achieved.}


\maketitle

\section{Introduction}\label{s:intro}
Time-variable accretion is commonly observed in low-mass pre-main-sequence stars, which results in strong and
long-lived optical bursts that are clearly visible in FU~Orionis-type objects \citep{audard2014}. In high-mass young stellar
objects (HMYSOs), similar phenomena are rarely observed because of their relatively low numbers, far distances, and strong extinction even at near-infrared wavelengths. Recent numerical simulations have predicted luminous bursts in
high-mass star formation induced by episodic accretion due to disk fragmentation \citep{meyer2017}.
These works showed that a rapid increase in the accretion rate from $10^{-3}$ to $10^{-1}$M$_{\odot}$yr$^{-1}$,
caused by a 0.55M$_{\odot}$ gaseous clump falling over $\sim$10\,yr from the circumstellar disk onto the protostar,
produces a luminosity burst of $\ge5\times10^5$L$_{\odot}$.

The singular evolution of the short-lived ($\sim$30\,yr) outflow traced by 22\,GHz water masers in the HMYSO W75N(B)-VLA~2
was probably related to an episodic increase in the accretion rate (\citealt{carrasco2015}).
Remarkable signatures of rapid accretion episodes have recently been reported in NGC6334I-MM1 \citep{hunter2018}
and S255IR-NIRS3 \citep{carratti2017}. In the first HMYSO, imaged in (sub)millimeter continuum, the luminosity
increased by a factor of $\sim$70 between 2008 and 2015/2016 and unprecedented maser activities in OH, H$_2$O,
and CH$_3$OH transitions have been discovered. The second object experienced a flare in the 6.7\,GHz methanol
maser line; the flux density increased by a factor of ten over $\sim$100~days \citep{fujisawa2015}.
Subsequent near-infrared observations revealed an increase in brightness by $\sim$3.5 and $\sim$2.5~mag
at H and K bands, respectively. Furthermore, spectral features of CO band-heads, NaI, and HeI lines appeared,
while the H$_2$ and Br$\gamma$ infrared lines increased by a factor greater than 20. These latter emissions are
typical signposts of accretion and ejection activity \citep{carratti2017}. High angular resolution observations uncovered
an extended (0\farcs2-0\farcs3) plateau of 6.7\,GHz maser emission newly created and likely associated with the increase 
in the infrared luminosity due to enhanced accretion  \citep{moscadelli2017}.

The source S255IR-NIRS3, also known as S255IR-SMA1 or G192.600$-$0.048, is being intensively studied in both line and continuum emissions
(e.g., \citealt{wang2011}; \citealt{zinchenko2015}). It is a $\sim$20\,M$_{\odot}$ protostar with a bolometric luminosity of 
$\sim$2.4$\times$10$^4$L$_{\odot}$ (\citealt{wang2011}; \citealt{zinchenko2012, zinchenko2015}) at the trigonometric distance of 
$1.78\pm0.12$\,kpc \citep{burns2016}. It exhibits a rotating disk-like structure with nearly edge-on appearance in 
the mid-infrared \citep{boley2013}. A large-scale ($\sim1$\arcmin) bipolar CO outflow with a redshifted lobe at position 
angle 67\degr\, is nearly perpendicular to the disk axis (\citealt{zinchenko2015}) and indicates at least three major ejection 
episodes with time intervals of 6000 and 800 years. The 22\,GHz maser data suggest that the third ejection
occurred $\le$130\,yr ago (\citealt{burns2016}). The motion of the light echo observed in the near-infrared implies
that the recent burst began in June 2015 (\citealt{carratti2017}).

In this paper we report long-term, high-cadence, single-dish observations of S255IR-NIRS3 in the 6.7\,GHz methanol
maser transition, which provide the timing of episodic accretion events complementary to the high angular resolution data
at radio and infrared wavelengths.

\section{Observations}\label{s:data}
The target was monitored from February 2013 to January 2018 with the Torun 32\,m radio telescope as part of the long-term program 
described in \citet{szymczak2018}. The average observation cadence was 6.1 per month, except for a few gaps of
1$-$2.5 months that were due to scheduling constraints. The 6668.519\,MHz methanol maser transition spectra were obtained with a dual-channel
HEMT receiver along with a $2^{14}$-channel autocorrelation spectrometer. The typical system equivalent flux density was $\sim$600\,Jy
before May 2015 and $\sim$450\,Jy afterward. The flux density scale was determined by observing the continuum source 3C123 and
the maser source G32.744$-$0.076 that contains spectral features that are non-variable within 5\%. The flux density calibration
uncertainty was about 10\%.
The back-end was configured to yield 0.09\,\kms\, spectral resolution after Hanning smoothing. Signals taken in frequency-switching
mode in the two orthogonal polarizations were averaged, and the typical rms noise level was 0.35\,Jy before 2015 May and 0.25\,Jy
afterward.

 \begin{figure}
  \centering
  \includegraphics[width=1.0\linewidth]{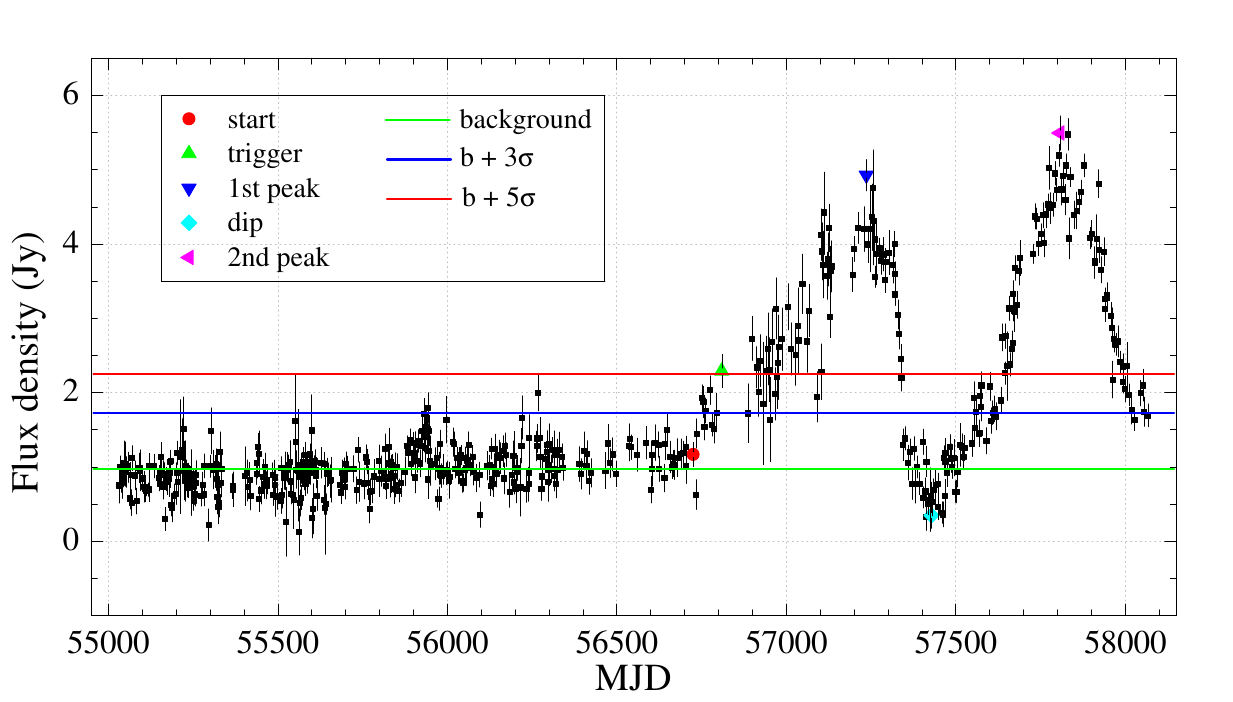}
  \caption{Typical  6.7\,GHz light curve for the 2.87\,\kms spectral channel of S255-NIRS3. The green line denotes the estimated
           background (b) flux level, while the blue and red lines indicate variations above the background by 3$\sigma$ and 5$\sigma$,
           respectively. The symbols in the legend indicate the start of the burst, the data point triggering the burst, the first and second peaks of
           the burst, and the dip of the light curve.
 \label{example-lc}}
 \end{figure}

In order to characterize the light curve for each individual spectral channel, we followed the thresholding approach of \cite{pietka2017}.
The background flux density level, {\it b},  and the signal variation due to noise, $\sigma$, were estimated from the pre-burst part
of the light curve, that is, before MJD $\sim$56716. This is a simple task because in this period the 6.7\,GHz emission generally displayed little or no variation.
The algorithm identifies data points where the flux density is more than 5$\sigma$ above the background level, so-called trigger points. When a trigger is found,
the routine searches for the nearest data points before and after the trigger with flux densities lower than 1$\sigma$ above the background
flux density: these are designated as the burst start and end, respectively. Figure~\ref{example-lc} illustrates  a typical light curve showing
the start, trigger, dip, and peaks of the burst markers. We do not show the end of the burst because the target is still in decaying phase
for most of the maser features.

We complemented the single-dish data with the recent VLBI observations reported in \cite{moscadelli2017}. We reduced the data of their experiment (code: RS002)
carried out on 2016 April 12 (MJD 57490) in phase-reference mode using 0603$+$1742 as phase calibrator. We employed standard procedures for spectral
line data (e.g., \citealt{bartkiewicz2016}). A typical rms noise level in line-free channels of 7\,mJy\,beam$^{-1}$ was achieved with the beam of
7.8$\times$4.9\,mas at position angle $-$26\fdg6.

\section{Results and analysis}\label{s:results}
\subsection{Total light curve}
  \label{sec:tlc}
\begin{figure}
   \centering
   \includegraphics[width=1.0\linewidth]{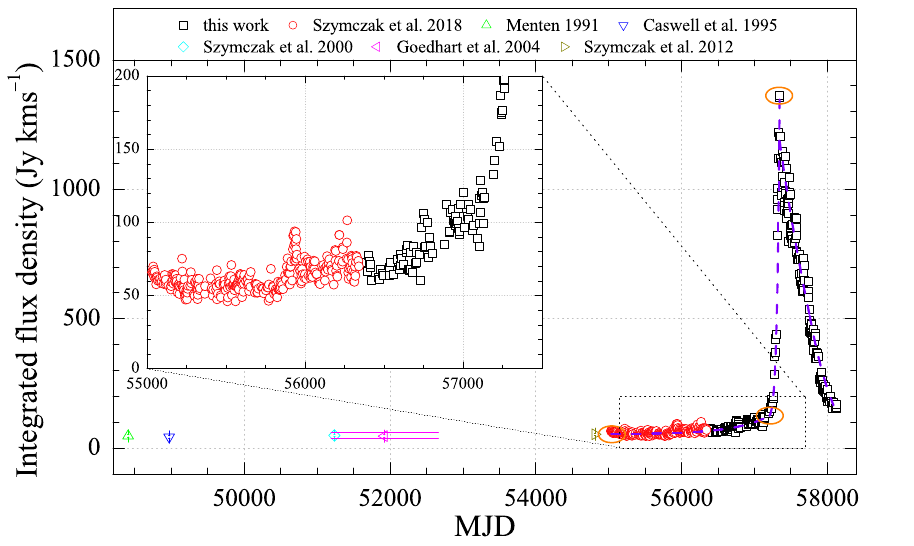}
   \caption{6.7\,GHz light curve of S255IR-NIRS3 spanning $\sim$27~years. The integrated flux density from different observations is displayed in differently
            colored symbols depicted in the legend. The data from \cite{goedhart2004} are represented by the average integrated flux density
            centered at the monitoring period of $\sim$1450~days shown by the two horizontal magenta bars; the top and bottom bars correspond
            to the upper and lower values of the integrated flux density, respectively. The dashed violet line indicates the result of fitting 
            three exponential functions to the data in different time intervals; the orange ellipses mark the starting points of each different fit interval.
            The inset shows the enlarged part of the light curve where an exponential rise starts.
   \label{total-lc}}
   \end{figure}

The long-term 6.7\,GHz light curve is shown in Fig.~\ref{total-lc}. The velocity-integrated flux densities, $S_{\mathrm{int}}$, from our
observations are complemented with the data available in the literature (\citealt{menten1991}; \citealt{caswell1995}; \citealt{goedhart2004};
\citealt{szymczak2000, szymczak2012, szymczak2018}). From $\sim$51190 to 52695, the maser showed a modest relative amplitude variation of 0.61, 
and the average $S_{\mathrm{int}}$ of 46.5\Jykms\, derived from the observations of \cite{goedhart2004} lies well within the range of 43$-$53\,\Jykms\, 
reported between roughly 48408$<$MJD$<$54811 (\citealt{menten1991}; \citealt{caswell1995}; \citealt{szymczak2000, szymczak2012}). 
During the period MJD~55033$-$55810, $S_{\mathrm{int}}$ was 57.7$\pm$5.1\,\Jykms\, (\citealt{szymczak2018}), then increased to 69.5$\pm$5.1\,\Jykms\, 
from MJD 55811 to 56716 with a 91\,\Jykms\, peak around MJD 55930. We argue that the available data indicate a quiescent state over 23\,years 
since the source was discovered; the emission showed only moderate variability less than 25-30\% on timescales of months to years.

Since MJD 56716, the maser exponentially increased from 79\,\Jykms\, to 1362\,\Jykms\, at MJD 57339; a particularly rapid increase in
$S_{\mathrm{int}}$ by a factor of 10 occurred in the MJD range of 57195$-$57339, then the emission started to decline exponentially.
The rising phase of the burst can be fitted by two exponential functions with characteristic times of 486 and 34 days,
while the decaying phase is fitted by one exponential law (Table~\ref{table1}) that is less accurate due to significant fluctuations
of $S_{\mathrm{int}}$. Very recent observations (January 2018), not presented here, show that the exponential decline is still continuing.

  \begin{table}
    \caption{Best-fit characteristic times of rise ($\tau_{\mathrm{r}}$) and decay ($\tau_{\mathrm{d}}$) of the burst for the integrated flux density.}
    \label{table1}
    \centering
    \begin{tabular}{l c c c }
    \hline
     MJD interval &  $S_{\mathrm{int}}$ range & $\tau_{\mathrm{r}}$   & $\tau_{\mathrm{d}}$ \\
                 &  (\Jykms)  &        (day)               &    (day)                 \\
    \hline
    55225-57195  &  55.6; 132.6  & 486.$\pm$41.5       &                     \\
    57195-57339  & 132.6; 1362   & 34.4$\pm$1.0       &                     \\
    57339-58075  & 1362; 165.4   &                    & 340.$\pm$68.        \\
    \hline
    \end{tabular}
   \end{table}

\subsection{Light curves of individual features}\label{sec:feat-profiles}
The dynamic spectrum in Fig.~\ref{dyn-spectrum} depicts a concise summary of the 6.7\,GHz maser intensity variations
over the 8.5~years when the super-burst occurred. The time series of the flux density integrated over selected velocity intervals
are plotted in Fig.~\ref{sel-LC-int}. These velocity intervals broadly correspond to the widths of individual spectral features
and are chosen to visualize the substantial variations of maser intensity.

\subsubsection{Pre-burst phase}
From MJD 55032 to $\sim$56716, the bulk of emission ranged from 3.6 to 6.0\,\kms\, and consisted of three persistent features 
at 4.10, 4.63 and 5.47\,\kms\, with a variability less than 25-30\%. A faint or marginally detected emission was present in
the velocity range of 1.65 to 3.60\,\kms, and the 1.82\,\kms\, feature showed long ($\sim$250-340~days) and faint (2.0-2.5\,Jy) bursts around
MJD 55524 and 56192. Since MJD 56240, a feature near 2.40\,\kms\, appeared, while a weak (1.9\,Jy) feature near 3.05\,\kms\, peaked around MJD 55936.

\subsubsection{Burst phase}
At velocities lower than 3.8\,\kms, new faint spectral features appeared after MJD $\sim$56716, exhibiting a variability pattern 
similar to that for the persistent features: the intensity increased over a period of about 600\,days, then rapidly (over 30-35\,days) 
decreased below the pre-burst level, and after $\sim$120\,days slowly increased again, reaching a second maximum around MJD 57700 (Fig.~\ref{sel-LC-int}).
The intensity of features in the velocity range of 4.8 to 5.6\,\kms\, rapidly increased around MJD 57300, just when the blueshifted
($<$3.8\,\kms) emission declined fast, by a factor of up to 8 relative to the pre-burst level, then decreased exponentially.
New emission features at velocities higher than 6.1\,\kms\, were detected after MJD 56800 until MJD 57280, they peaked around MJD 57315,
and then exponentially disappeared. 
A detailed description of the time variation in flux density of individual maser features during the burst is given in Appendix A.
                    
 \begin{figure}
  \includegraphics[width=1.0\linewidth]{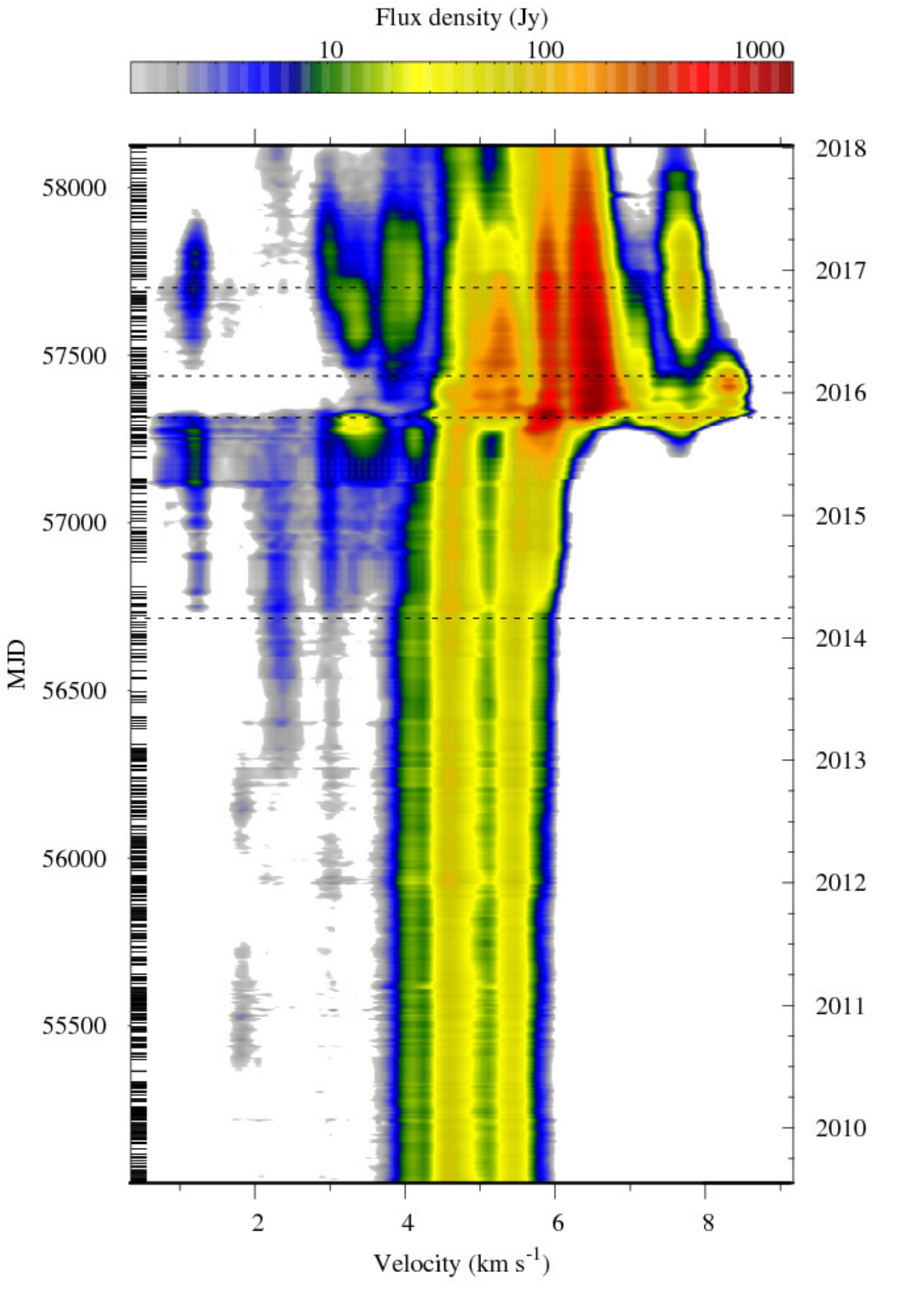}
  \caption{Dynamic spectrum of the 6.7\,GHz methanol maser emission for S255IR-NIRS3. The color scale maps to the flux density as shown
           in the wedge on the top. The flux densities are linearly interpolated between consecutive 32\,m telescope spectra.
           The velocity scale is relative to the local standard of rest. Individual observation dates are indicated by black tick marks
           near the left ordinates. The horizontal dashed lines from bottom to top mark the approximate times of the start, first peak, 
           dip, and second peak of the burst.
  \label{dyn-spectrum}}
  \end{figure}

 \begin{figure*}
 \centering
 \includegraphics[width=1.0\linewidth]{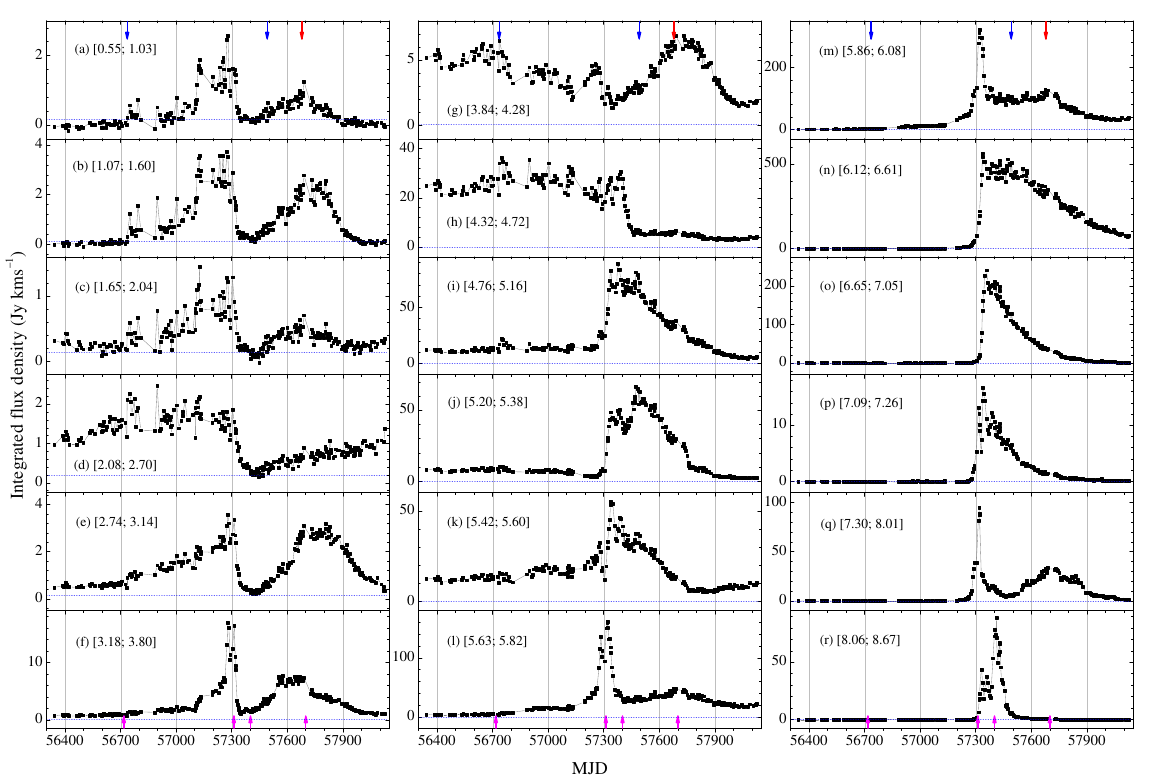}
 \caption{Burst profiles of the 6.7\,GHz methanol emission. The time series of the integrated flux density for 18 distinct velocity intervals are shown.
          The velocity intervals, given between square brackets at the top left of each panel, are selected in order to visualize the substantial changes in maser intensity
          and largely correspond to the widths of individual spectral features. The dotted line in each panel marks the 1$\sigma$ level. The consecutive magenta
          arrows above the bottom abscissa denote the approximate times of the start, (MJD $\sim$56716), the first peak (MJD $\sim$57310), the dip (MJD $\sim$57400),
          and the second peak (MJD $\sim$57700) of the burst. The red and blue arrows show the dates of JVLA (see Fig.~\ref{s255-vla}) and VLBI
          (see Fig.~\ref{s255-evn}) observations, respectively.
  \label{sel-LC-int}}
  \end{figure*}

\subsection{Overall characteristics}
The light curves of individual features are quite different, suggesting that either spatially separated regions respond differently
to the same exciting pulse or that their physical conditions vary significantly in time. However, there is a general trend of suppression
of the maser emission in the blueshifted part of the spectrum ($<$4.7\,\kms) when the redshifted emission ($>$4.7\,\kms) grows rapidly. 
The clear dip in the light curves of the blueshifted emission synchronously occurred with the appearance of new emission in
the velocity ranges of 4.76-5.60, 6.12-7.26, and 8.06-8.67\,\kms. The emission in these velocity intervals is generally characterized
by a rapid growth and exponential decline, while the burst profile of the emission in the ranges of 5.63-6.08 and 7.30-8.01\,\kms\, is very similar
to that of the blueshifted part of the spectrum, where a dip is clearly seen between the first and second peaks.

 \begin{figure}
 \centering
 \includegraphics[width=1.0\linewidth]{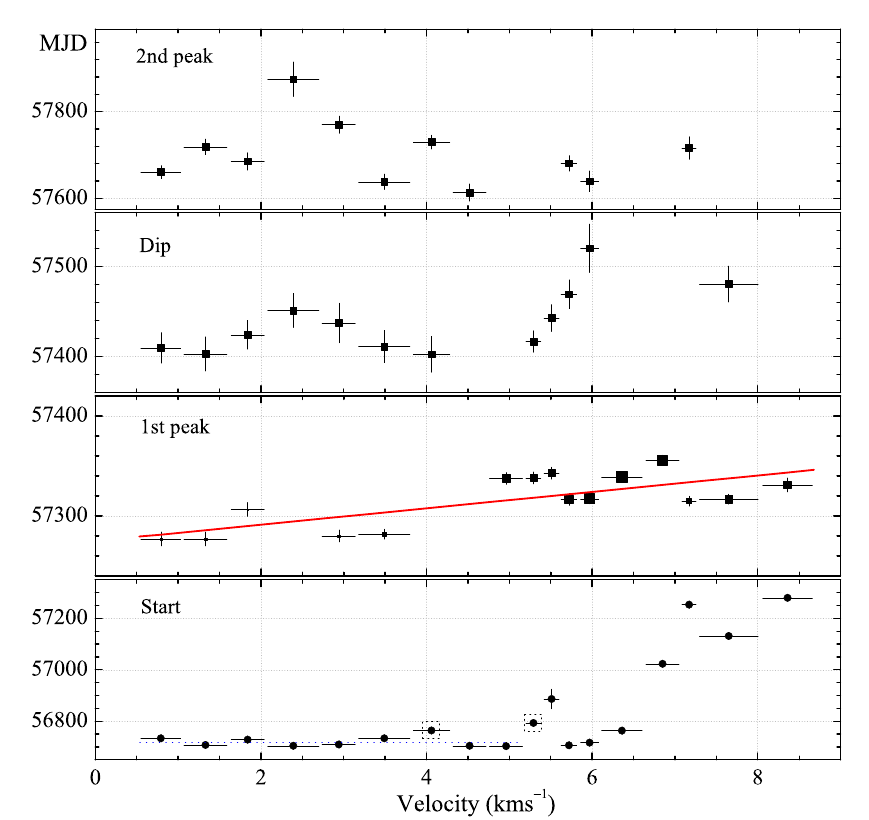}
 \caption{Timing characteristics of the burst vs. radial velocity. Shown are the times of start, first peak, dip, and second peak obtained by
          averaging the light curves over the same velocity intervals as in Fig.~\ref{sel-LC-int}. All the horizontal bars denote the widths of these velocity intervals.
          Bottom panel: Errors for the start time are smaller than or comparable to the symbol size, the dotted line indicates the average 
          time of the burst onset for the persistent emission, and the two data points surrounded by dotted square denote the beginning
          of decline in intensity. Second (from bottom) panel: Symbol sizes are proportional to the logarithm
          of the peak-integrated flux density, and the red line shows the linear fit to the time of the first peak vs. velocity.
  \label{time-summary}}
  \end{figure}

Fig.~\ref{time-summary} depicts the timing characteristics of the burst. For the emission at velocities lower than 5.2\,\kms, the burst
started on MJD 56716$\pm$5. In this estimate we excluded the velocity interval 3.84-4.38\,\kms\, because only the start of a slight decay
could be determined. The same applies to the emission at 5.20-5.38\,\kms, and both intervals are marked with dotted squares (Fig.~\ref{time-summary}).
The onset of the burst at 5.42-5.60\,\kms\, is poorly estimated as the course of the flux density was very flat before MJD 56809 and no data are
available for the consecutive $\sim$2.5~months. The appearance of the emission in the velocity ranges of 6.12-6.61\,\kms\, and higher than 6.65\,\kms\,
was delayed by 1.5 and 10-18.5~months, respectively.

The emission peaked on MJD 57315$\pm$11, and the flux-weighted mean epoch of the peak was MJD 57334$\pm$10 (Fig.~\ref{time-summary}).
In the velocity range 3.84 to 4.72\,\kms, the time of the peak could not be determined. The estimated mean time of the peak was MJD 57284$\pm$6 and
57331$\pm$10 for the emission in the velocity ranges of 0.55-3.80 and 4.76-8.67\,\kms, respectively. A linear fit to the data shows a strong correlation
(r=0.738, p<0.0017) between the time of peak and velocity; the time delay between the features at the outermost extreme velocities is 62$\pm$16\,days.

At 47-195\,days (with mean and dispersion of 142 and 12\,days, respectively) after the maximum, the burst profile attained a dip at a pre-burst or
lower level for the emission at velocities $<$4.28\,\kms\, and at higher than a pre-flare level for the new emission at velocities $>$5.6\,\kms.
The mean time of the dip was MJD 57439$\pm$10. The second peak occurred on MJD 57702$\pm$22, that is, 406$\pm$30 days after the first peak.

The burst lasted 324$\pm$36, 707$\pm$19 and 1264$\pm$8 days for the emission in the velocity ranges of 8.06-8.67, 7.09-7.26 and 1.07-1.60\,\kms, respectively.
For most of the features, the emission is still decaying, and the lower limit of the burst duration is 1410\,days.

\begin{figure}
 \includegraphics[width=1.0\linewidth]{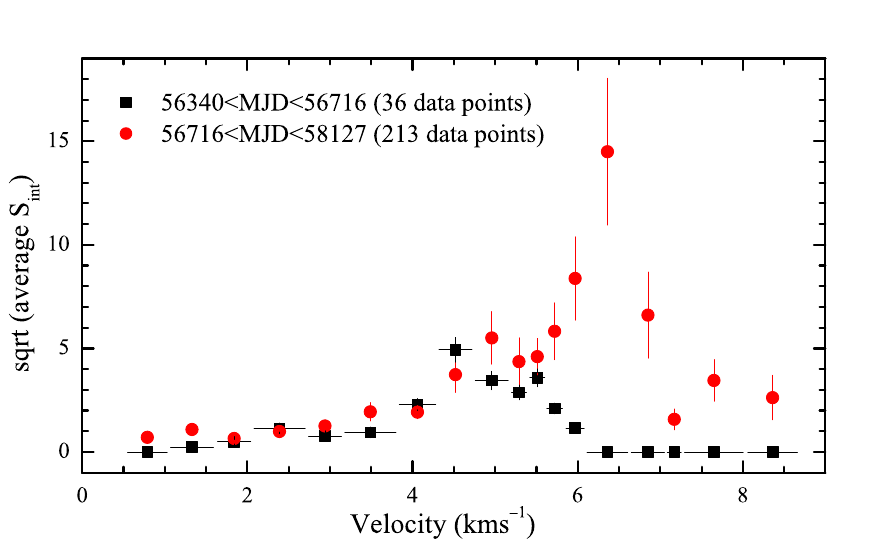}
 \caption{Integrated flux density ($S_{\mathrm{int}}$) before (black squares) and during the burst (red circles).
          The root square of mean $S_{\mathrm{int}}$ over the same velocity intervals as in Fig.~\ref{sel-LC-int} is shown.
          The time spans and the number of data points are given in the legend.
 \label{s225-averS}}
 \end{figure}

Fig.~\ref{s225-averS} shows that the  main ($\sim$79\%) contribution to the integrated flux density in the burst comes from new features at velocities
$>$5.8\,\kms. It is clearly visible that in the velocity intervals 2.08-2.70 and 3.84-4.72\,\kms, the emission is suppressed by 14-25\%
during the burst.  

During the rising phase and around the peak of the burst (57200$<$MJD$<$57500), the emission in all spectral channels showed significant modulations of
20 to 75\% in amplitude on timescales of 10 to 35~days with an average of 23.8$\pm$7.4~days. These modulations are often
synchronized within a few days for individual features and may reflect temporal variations of the underlying pulse of radiation
that triggers the maser burst or are signposts of non-equilibrium conditions in the maser regions.

During the burst, several spectral features displayed drifts in peak velocity (Fig.~\ref{dyn-spectrum}). For instance, the peak velocity of the strongest feature
changed from 6.38 to 6.56\,\kms\, over 38\,days (57311$<$MJD$<$57349) and returned to the initial value after 648\,days. These drifts can be
most naturally explained in terms of excitation of gas layers that differ only slightly in radial velocity.

\section{Discussion}\label{s:discussion}
In our study, the long-term and high-cadence monitoring of the 6.7\,GHz line clearly shows an extraordinary
burst in S255IR-IRS3 triggered by episodic accretion (\citealt{carratti2017}). This is a prominent source in the very small group of HMYSOs
that shows signatures of luminous accretion events (\citealt{carrasco2015}; \citealt{hunter2018}), providing a unique opportunity for exploring
the burst mechanism.

\subsection{Burst onset and duration}
It is striking that the rapid increase in the 6.7\,GHz maser integrated flux density that started on MJD 57195 perfectly coincides with
the onset of the burst (mid-June 2015, i.e., MJD 57188) estimated from the motion of the light echo imaged in Ks/K band (\citealt{carratti2017}).
Therefore, a casual relationship is clear among 1) the infrared brightening of the central source and its outflow cavities, 
2) the appearance of emission lines typical of accretion bursts in eruptive low-mass stars (but from three to four orders of magnitude stronger)
(\citealt{carratti2017}), and 3) the maser burst. The burst of radio continuum at 6-46\,GHz that started after MJD 57579 has confirmed
the accretion interpretation of the maser burst in this source (\citealt{cesaroni2018}).

Hydrodynamic simulations of massive star formation predict episodic luminosity bursts induced by accretion of gaseous clumps falling
from the circumstellar disk onto the protostar (\citealt{meyer2017}). These models produce spikes of total luminosity from $\sim1.5\times10^4$L$_{\odot}$
to $\sim5\times10^5$L$_{\odot}$ over a time interval of 10\,yr induced by a rapid increase in the accretion rate from $\sim3\times10^{-3}$M$_{\odot}$yr$^{-1}$
to $\sim10^{-1}$M$_{\odot}$yr$^{-1}$. Figure~\ref{total-lc} implies that the maser flare lasted longer than $\sim$2.5\,yr and was preceded by
a moderate increase in flux density in the blueshifted ($<$3.8\,\kms) part of the spectrum over a time span of $\sim$1.5\,yr.
The duration of the flare estimated for the blueshifted features for which the flux density returned to the pre-burst levels (Fig.~\ref{sel-LC-int}a, b, e, f) is $\sim$3.5\,yr.
We conclude that the duration of the maser burst is comparable within a factor of three with that of the luminosity burst in the numerical models
by \cite{meyer2017}. The model burst is preceded by a slow ($\sim$40\,yr) increase in the accretion rate of $2\times10^{-3}$M$_{\odot}$yr$^{-1}$ to
$7\times10^{-3}$M$_{\odot}$yr$^{-1}$ , which results in a rise of the total luminosity from $10^4$L$_{\odot}$ to $2.5\times10^4$L$_{\odot}$.
Fig.~\ref{total-lc} does not show evidence for significant changes of the maser-integrated flux density over $\sim$23\,yr before the burst.
                
\subsection{Energy conversion efficiency}
It is thought that $\sim$20-30$\mu$m photons are required in the 6.7\,GHz methanol pumping scheme (\citealt{sobolev1997}). Using the infrared measurements
of \cite{carratti2017} (their Table 1 and Fig. 3) and the methanol maser data, we attempt to evaluate the infrared-to-maser energy
conversion efficiency in the quiescent and burst periods in a very simple manner. Assuming that the maser and infrared photons are produced in the same volume of gas,
we obtain  $\eta_{\mathrm{19.7}}=S_{\mathrm{int}}/(S_{\mathrm{19.7}}\Delta V_{19.7})$
equal to $0.07\pm0.02$ and $0.21\pm0.03$ in the pre-burst period and in February 2016 (MJD 57422), respectively. The corresponding values of
$\eta_{\mathrm{31.5}}=S_{\mathrm{int}}/(S_{\mathrm{31.5}}\Delta V_{31.5})$ are $0.043\pm0.010$ and $0.039\pm0.010$. Here, $S_{\mathrm{19.7}}$
and $S_{\mathrm{31.5}}$ are the flux densities at 19.7 and 31.5$\mu$m, respectively. $\Delta V_{19.7}$=$\Delta V_{31.5}$ is the range of radial
velocity across the methanol maser spectrum, which is 4.0 and 8.1\,\kms\, for the quiescent and burst periods, respectively.
These crudely estimated ratios suggest that the conversion efficiency of 19.7$\mu$m emission into the maser emission
increases by a factor of two during the burst, whereas at 31.5$\mu$m, it does not differ significantly between the pre-burst and burst periods.
We assume that the efficiency at 19.7$\mu$m increases during the burst because the maser amplification is likely closer to saturation, as the high 
maser intensity suggests (\citealt{moscadelli2017}).

The star luminosities in the H (1.65$\mu$m) and Ks (2.16$\mu$m) bands in the burst exceed the pre-burst values
by a factor of 25 and 10, respectively (\citealt{carratti2017}). The total star luminosity estimated from
spectral energy distribution increased by up to a factor of 6.  Fig.~\ref{total-lc} indicates that the methanol maser was in a quiescent state 
over $\sim$23 years with a mean luminosity of ($1.2\pm0.3)\times10^{-6}$L$_{\odot}$ , and during the burst, it increased by a factor 18.
                                     
\subsection{Changes in maser morphology and luminosity}
\begin{figure}
 \centering
 \includegraphics[width=1.05\linewidth]{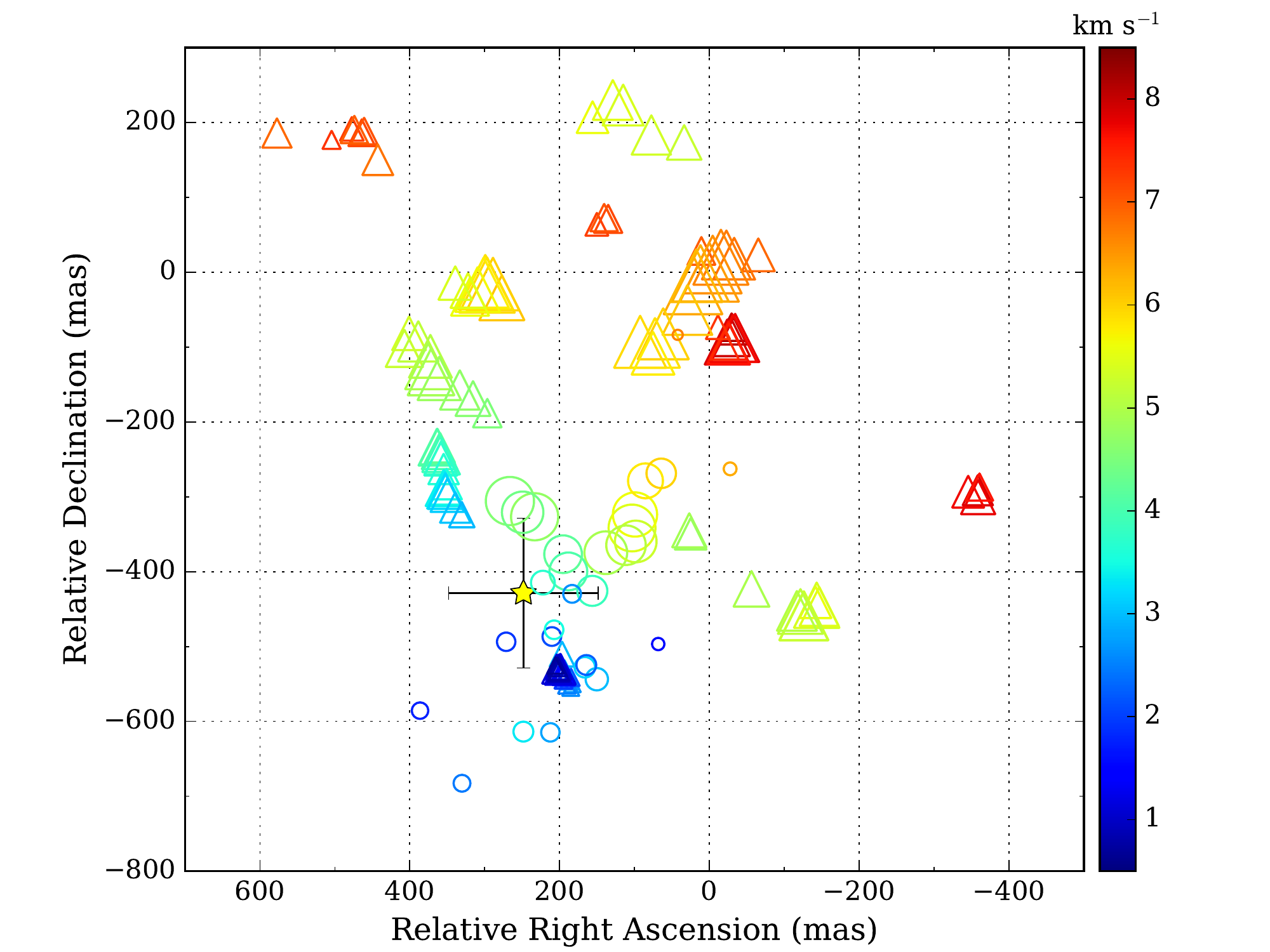}
 \caption{Distribution of the 6.7\,GHz methanol maser spots in S255IR~IRS3 taken before (circles) and after (triangles) the burst on MJD 55996 (JVLA-C, \citealt{hu2016})
          and MJD 57676 (JVLA-A, \citealt{moscadelli2017}), respectively. The origin of the map coincides with the brightest spot at $\alpha$(J2000) = $06^{\mathrm{h}}12^{\mathrm{m}}53\fs998$,
          $\delta$(J2000)= 17\degr59\arcmin23\farcs49. The relative position uncertainty of the spots between these two epochs is less than $\sim$0\farcs1.
          The symbol size is proportional to the maser brightness, and colors denote the radial velocity in \kms\, according to the right scale.
          The cross marks the position of the 5\,GHz continuum emission (\citealt{moscadelli2017}).
 \label{s255-vla}}
 \end{figure}

\begin{figure}
 \centering
 \includegraphics[width=1.0\linewidth]{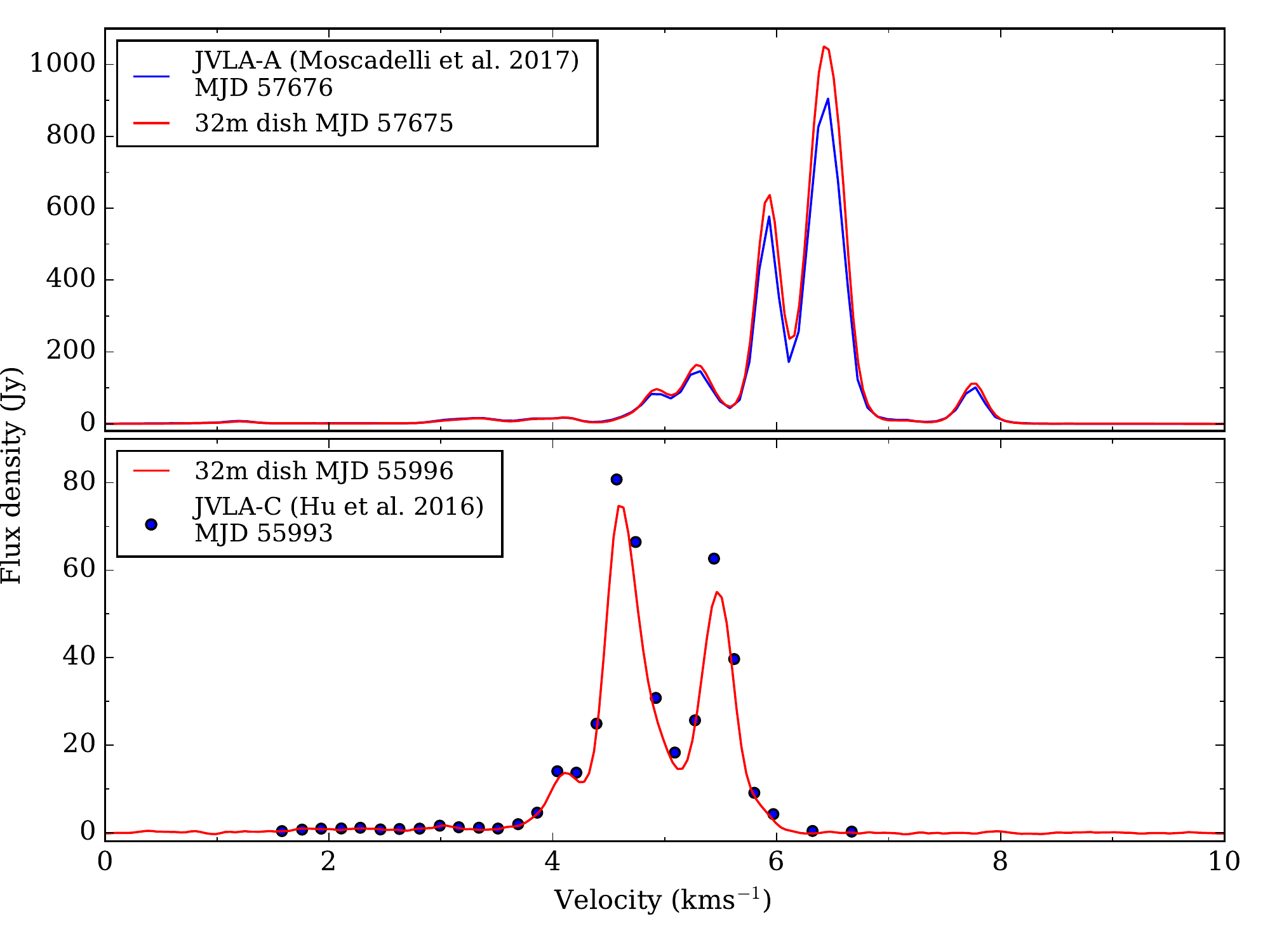}
 \caption{Spectra at 6.7\,GHz before and after the burst obtained with the JVLA and 32\,m telescope. The flux density differences between the nearly simultaneous spectra are well
          within the calibration accuracy of the 32\,m dish.
  \label{s255-spec-comp}}
  \end{figure}

The target was observed before the burst on MJD 55993 using the JVLA in C-configuration (\citealt{hu2016}), and 
it appears that the 6.7\,GHz maser emission had roughly ellipsoidal ($\sim0\farcs7\times0\farcs3$)
distribution with an overall area of $\sim$0.16 square arcseconds and a velocity gradient along the major axis at a position angle (PA) of $-31\degr$
(Fig.~\ref{s255-vla}). This major axis is nearly parallel to the major axis (PA$\approx-27\degr$) of the rotating disk
seen in the thermal lines of CH$_3$OH and CH$_3$CN at 220\,GHz (\citealt{zinchenko2015}). However, the beam of this JVLA observation was about 
4\arcsec , and we cannot derive any firm conclusion on the maser spatial distribution on scales of $\sim0\farcs5$.
We can only state that this JVLA-C observation fully recovered the entire 32\,m dish flux density
(Fig.~\ref{s255-spec-comp}), and the maser luminosity was $(1.4\pm0.1)\times10^{-6}$L$_{\odot}$.

Nearly one year after the burst peak, the JVLA observation on MJD 57676 in A-configuration (\citealt{moscadelli2017} their Fig.~3) revealed
a maser emission with an exceptionally extended morphology (Fig.~\ref{s255-vla}).  The major axis of the emission structure was $\sim1\farcs1$ at PA $\approx44\degr$.
The extent of the emission along the plane of disk at PA $=-27\pm4\degr$ (\citealt{boley2013}; \citealt{zinchenko2015}) was $\sim$0\farcs7.
The total area of the maser region increased to $\sim$0.6 square arcseconds, while the maser luminosity was $(1.4\pm0.1)\times10^{-5}$L$_{\odot}$ at MJD 57676.
The size of the maser region is probably underestimated because at these after-burst epochs and extended JVLA configurations, about 10-15\% of the single-dish flux density
was missed on the longest baselines (\citealt{moscadelli2017}). It is striking that the shape of the maser region was very similar to that observed in 0.85\,mm
dust emission; in particular, a sharp drop in intensity at the eastern side of the disk and extended lobe in the N-W direction were seen (\citealt{zinchenko2017}).

\begin{figure}
 \centering
 \includegraphics[width=1.0\linewidth]{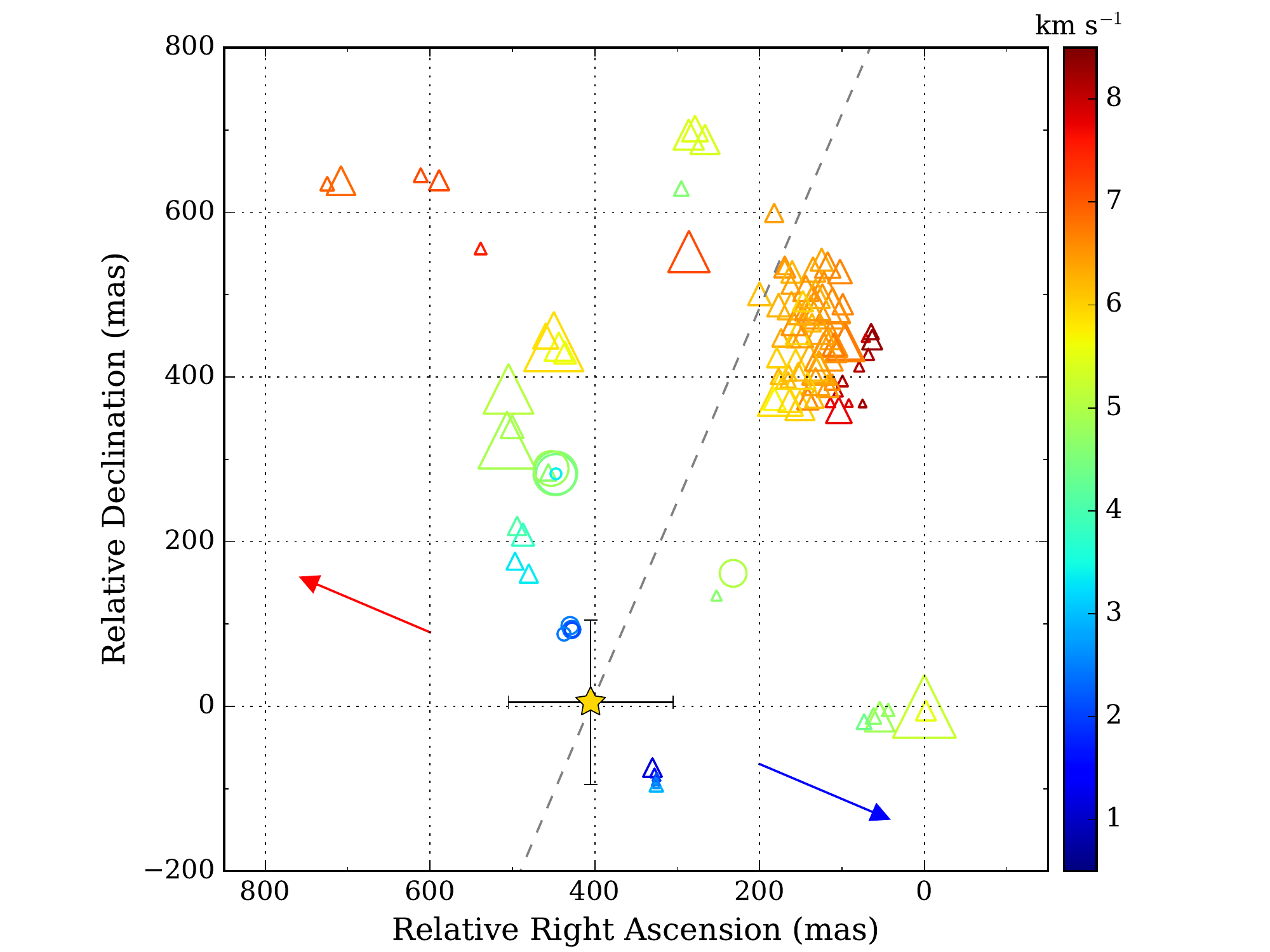}
 \caption{Distribution of the 6.7\,GHz methanol maser spots in S255IR~IRS3 taken on MJD 56736 (circles, http://bessel.vlbi-astrometry.org/data) and MJD 57490 (triangles, \citealt{moscadelli2017})
          using the VLBA and EVN, respectively. The relative position uncertainty of the spots between these two epochs is less than a few tens of mas. The symbol size is proportional to the maser brightness, and
          colors denote the radial velocity in \kms\, according to the right scale. The cross marks the position of the 5\,GHz continuum emission (\citealt{moscadelli2017}).
          The dotted gray line marks the major axis of the disk, whereas the arrows indicates the blue- and redshifted lobes of large-scale CO outflow (\citealt{zinchenko2015}).
 \label{s255-evn}}
 \end{figure}

Figure~\ref{s255-evn} shows the distribution of 6.7\,GHz maser spots observed with VLBI 1.54\,yr before and 0.52\,yr after the burst peak. It must be noted
that the maser emission was highly resolved out with VLBI beams of $\sim5\times3$mas. In the pre-burst observation on MJD 56736, the VLBA correlated flux density
of the strongest feature at 4.64\,\kms\, was only 5.5\% of that observed with the 32\,m dish; the ratio of $S_{\mathrm{int}}$ at VLBA and 32\,m dish was only 0.024.
The EVN observation on MJD 57490 recovered less than 10\% of the entire flux density measured with the 32\,m dish. Therefore, before and after the burst peak,
the majority of the emission comes from extended low-brightness structures.
Nevertheless, the VLBI data provide evidence of huge transformations of the maser morphology; the blueshifted ($<$5\,\kms) spots observed before
the burst peak close to the radio continuum source completely disappeared, and after the burst peak, emission at similar blueshifted velocities rose from outer layers
displaced by 110-170\,mas from the pre-burst positions; new emission at $>$5\,\kms\, appeared 340-670\,mas from the continuum source. For the adopted distance
of 1.78\,kpc, these angular offsets correspond to a projected distance of 200-300 and 600-1200\,au, respectively.
This suggests that the accretion burst illuminates more distant layers of the gas, and low-brightness maser emission arises from extended area.
                                     
The maser emission observed with the EVN about 16.4\,yr before the burst peak can be divided into four distinct clusters (\citealt{moscadelli2017} their Fig. 2)
on the basis of the similar spatial distribution and radial velocities. Three of these groups were still present after $\sim$15\,yr when observed with the VLBA (Fig.~\ref{s255-evn}),
although the morphology of individual groups, especially the western one, differed significantly; the southern group at blueshifted velocities was not detected.
These differences are likely due to long-term changes in the maser conditions and partly due to a spectral resolution lower by a factor of two and to the lower sensitivity of the VLBA
observation as compared to the EVN observation (\citealt{moscadelli2017}). The overall spatial structure of the most compact components and their radial velocities
remained stable before the burst peak.

Summarizing the results of VLBI observations, we conclude that during 4.6\,yr, the size of the maser region in S255IR-NIRS3 and its maser luminosity increased by
a factor of 3 and 10, respectively. The growth in the maser luminosity appears to be a simple consequence of the increasing volume of gas which reaches the maser inversion
conditions due to an increase in total luminosity caused by the accretion event. This is consistent with the statistical results showing that more luminous 6.7\,GHz methanol
maser sources are preferentially associated with dust clump sources that have higher radii and mass (\citealt{breen2010}).
The maser structure along the major axis of the disk increased in size by a factor of 3; the new red-shifted ($>$6\,\kms) maser components appeared
in layers at a projected distance higher than 1000\,au from the central star (\citealt{moscadelli2017}; \citealt{cesaroni2018}), while
the blueshifted ($<$3\,\kms) emission located close to the central star largely disappeared.

\subsection{Causes of the burst profile diversity}
Figures~\ref{dyn-spectrum} and \ref{sel-LC-int} depict large differences in the maser burst profiles of individual features, which suggests essential inhomogeneities in
the physical conditions of surrounding materials. The beam filling factor of 0.15-0.20 derived for the CH$_3$CN and CH$_3$OH thermal lines suggests a strongly
fragmented disk with small ($\ll$500\,au) clumps of density higher than the mean value of 3$\times10^8$\,cm$^{-3}$ and a gas kinetic temperature
of 180\,K (\citealt{zinchenko2015}). Such a clumpy medium supporting the maser amplification can produce a variety of response patterns to the accretion luminosity burst.

We note that the light curves of some features (Fig.~\ref{sel-LC-int}f, l, m) in S255IR-NIRS3 resemble that reported for feature $-$7.2\,\kms\,
in NGC~6334I (\citealt{hunter2017} their Fig.~4).  Its characteristic times of rise and decay of the burst are about 120 and 90\,days, respectively,
that is, a factor of 3-10 longer than for the above features in S255IR-NIRS3. On the other hand, the emission ranging from 2.7 to 3.1\,\kms (Fig.~\ref{sel-LC-int}e)
has a characteristic rise time (160\,days) of the burst comparable to that seen for the feature in NGC~6334I. This supports the hypothesis that the extraordinary
bursts of 6.7\,GHz masers in both sources were triggered by an accretion event.

The model calculations showed that the 6.7\,GHz maser operates in a wide range of gas density of $10^4-10^8$\,cm$^{-3}$ for a dust temperature higher than 150\,K
and a methanol fractional abundance of 10$^{-7.5}$ to 10$^{-5}$ (\citealt{sobolev1997}; \citealt{cragg2002}; \citealt{cragg2005}). The intensity diminishes when
the gas temperature approaches or exceeds the dust temperature, while the bright emission is predicted for the hydrogen density of 10$^5$ to 10$^8$\,cm$^{-3}$
and a methanol column density of 10$^{16}$ to 10$^{18}$\,cm$^{-2}$ (\citealt{cragg2002}). The following interpretation of the burst profiles largely relies on
this model.

In the first phase of the burst (56716$\la$MJD$\la$57300), a faint emission at velocities $\le$3.80\,\kms, coming probably from regions of density lower
than $10^7$\,cm$^{-3}$ (\citealt{zinchenko2015}), rises due to an increase in dust temperature that governs the distribution between the assorted torsional and vibrational states
of the molecule (\citealt{sobolev1997}). Much stronger persistent emission in the velocity ranges of 2.08-2.70 and 3.84-4.72\,\kms\, declines or remains stable
because it probably arises in higher density ($>10^{7}$\,cm$^{-3}$) regions where the kinetic temperature rises faster than in lower density regions
(e.g., \citealt{johnstone2013}). According to the maser model, the intensity decreases when the gas temperature exceeds the dust temperature (\citealt{cragg2005}).
Some features of the persistent emission from 4.76 to 6.08\,\kms\, decline or rise in intensity likely due to a complex transformation in the maser structure
and related changes in the maser optical depth that cause the disappearance of pre-burst maser clouds and appearance of new clouds (\citealt{moscadelli2017}).

After a rapid increase in the accretion luminosity around MJD 57300, the blueshifted emission ($<$3.80\,\kms) abruptly falls below the detection level
due to the destruction of the molecule or perturbations of the velocity coherent path. For a model of homogeneous density distribution for the molecular envelope
and isotropic radiation from the HMYSO, \cite{moscadelli2017} have shown that UV photodissociation of the molecule induced by the accretion burst plays
no significant role in the disappearance of this emission. Fig.~\ref{s255-evn} reveals that the pre-burst emission comes from regions
close ($\sim$340\,au) to the star, and the blueshifted emission could be strongly suppressed by the radiation field triggered by the accretion burst. Detailed calculations
by \cite{moscadelli2017} indicated that radiation pressure destroys the velocity coherence by 0.1-0.2\,\kms\, along a sizable fraction of the maser path length.
This effect, however, is an order of magnitude lower at a distance of $\sim$800\,au and does not affect the redshifted emission at velocities $\ge$4.8\,\kms.
The maser intensity rapidly increases with fast ($\sim$3 weeks) modulations, which also last around the burst peak probably because the dust and gas do not attain
temperature equilibrium (\citealt{johnstone2013}) in newly excited regions of the northern part of the envelope. The new redshifted ($>$6.12\,\kms) emission is
located at projected distances of 800-1000\,au from the central star (Figs.~\ref{s255-vla} and \ref{s255-evn}) and burst peaks of the redshifted features are delayed
by $\sim4\times10^6$\,s relative to the blueshifted features (Fig.~\ref{time-summary}). This delay is an order of magnitude longer than the photon travel time, but
is comparable to the gas heating time (\citealt{johnstone2013}), suggesting that the bright maser is turned on for a certain gas temperature.

During the decay phase, the blueshifted emission ($<$4.8\,\kms) resumes with a peak at $\sim$MJD 57700, but 
the VLBI maps imply that it comes from outside parts of the pre-burst maser region (Fig.\ref{s255-evn}) with the exception of the bursted feature centered at 1.2\,\kms.
It is possible that this resumed emission at new locations disappears when the velocity coherence is destroyed at a distance of more than 300\,au
due to a second peak in the luminosity burst. The light curves of all the features show more or less exposed humps around $\sim$MJD 57700,
which are evidence of the second peak of the burst.

\section{Conclusions}
We have monitored the 6.7\,GHz methanol maser in S255IR-IRS3 over a $\sim$5-year period and obtained the light curve during an extremely strong burst that was
likely induced by an episodic accretion event. The data were supplemented with our previously reported observations and those available in the literature.

We found that the maser luminosity in the burst peak was more than an order of magnitude higher than that during the 23-year quiescent period.
The maser emission underwent complex changes, including a strong suppression of the persistent blueshifted features just when the new emission appeared
in the redshifted part of the spectrum. This drastic transformation of the spectrum is probably due to perturbations of the velocity coherent path close
to the central star and the excitation of outer layers by the accretion luminosity burst. This interpretation is consistent with enormous changes 
in the maser morphology and an increase in its size by a factor of four. The onset of rapid increase in the maser intensity exactly coincides with that of the burst
estimated from the motion of the light echo observed in near-infrared emission, supporting a radiative pumping mechanism of the transition.

Further monitoring of S255IR-IRS3 would be suitable for single dishes and interferometric arrays in order to study the evolution of the accretion
burst and related phenomena.

\begin{acknowledgements}
We thank the Torun CfA staff and the students for assistance with the observations. We would like to thank Luca Moscadelli for kindly sharing
the JVLA maps with us and an anonymous referee for providing valuable suggestions to improve this paper.
This research has used the SIMBAD data base, operated at CDS (Strasbourg, France) and NASA's Astrophysics Data System Bibliographic Services.
The work was supported by the National Science Centre, Poland through grant 2016/21/B/ST9/01455.
\end{acknowledgements}

\bibliographystyle{aa}             
\bibliography{aa33443}       

\begin{thebibliography}{0}
\expandafter\ifx\csname natexlab\endcsname\relax\def\natexlab#1{#1}\fi



\bibitem[{{Audard} {et~al.}(2014){Audard}, {\'Abrah\'am}, {Dunham}, {Green}, {Grosso}, {Hamaguchi}, {Kastner}, {K\'osp\'al}, {Lodato}, {Romanova}, {Skinner}, {Vorobyov}, & {Zhu}}]{audard2014}
         {Audard}, M., {\'Abrah\'am}, P., {Dunham}, M.~M., {et~al.} 2014,  Protostars and Planets VI (Tucson: University of Arizona Press), 387

\bibitem[{{Bartkiewicz} {et~al.}(2016){Bartkiewicz}, {Szymczak}, & {van Langevelde}}]{bartkiewicz2016}
         {Bartkiewicz}, A., {Szymczak}, M., {van Langevelde}, H.~J. 2016, \aap, 587, A104

\bibitem[{{Boley} {et~al.}(2013){Boley}, {Linz}, {van Boekel}, {Henning}, {Feldt}, {Kaper}, {Leinert}, {M\"uller}, {Pascucci}, {Robberto}, {Stecklum}, {Waters}, & {Zinnecker}}]{boley2013}
         {Boley}, P.~A., {Linz}, H., {van Boekel}, R., {et~al.} 2013, \aap, 558, A24

\bibitem[{{Breen} {et~al.}(2010){Breen}, {Ellingsen}, {Caswell}, & {Lewis}}]{breen2010}
         {Breen}, S.~L., {Ellingsen}, S.~P., {Caswell}, J.~L., {Lewis}, B.~E. 2010, \mnras, 401, 2219

\bibitem[{{Burns} {et~al.}(2016){Burns}, {Handa}, {Nagayama}, {Sunada}, & {Omodaka}}]{burns2016}
         {Burns}, R.~A., {Handa}, T., {Nagayama}, T., {Sunada}, K., {Omodaka}, T. 2016, \mnras, 460, 283

\bibitem[{{Carratti o Garratti} {et~al.}(2017){Caratti o Garatti}, {Stecklum}, {Garcia Lopez}, {Eisl\"offel}, {Ray}, {Sanna}, {Cesaroni}, {Walmsley}, {Oudmaijer}, {de Wit}, {Moscadelli}, {Greiner}, {Krabbe},
         {Fischer}, {Klein}, & {Iba\~nez}}]{carratti2017} {Caratti o Garatti}, A., {Stecklum}, B., {Garcia Lopez}, R., {et~al.} 2017, Nature Physics, 13, 276

\bibitem[{{Carrasco-Gonz\'alez} {et~al.}(2015){Carrasco-Gonz\'alez}, {Torrelles}, {Cant\'o}, {Curiel}, {Surcis}, {Vlemmings}, {van Langevelde}, {Goddi}, {Anglada}, {Kim}, {Kim}, & {G\'omez}}]{carrasco2015}
         {Carrasco-Gonz\'alez}, C., {Torrelles}, J.~M., {Cant\'o}, J., {et~al.} 2015, Science, 348, 114

\bibitem[{{Caswell} {et~al.}(1995){Caswell}, {Vaile}, {Ellingsen}, {Whiteoak}, & {Norris}}]{caswell1995}
         {Caswell}, J.~L., {Vaile}, R.~A., {Ellingsen}, S.~P., {Whiteoak}, J.~B., {Norris}, R.~P. 1995, \mnras, 272, 96

\bibitem[{{Cesaroni} {et~al.}(2018){Cesaroni}, {Moscadelli}, {Neri}, {Sanna}, {Caratti o Garatti}, {Eisloffel}, {Stecklum}, {Ray}, & {Walmsley}}]{cesaroni2018}
         {Cesaroni}, R.~L., {Moscadelli}, L., {Neri}, R., {et~al.} 2018, \aap, 612, A103

\bibitem[{{Cragg} {et~al.}(2002){Cragg}, {Sobolev}, & {Godfrey}}]{cragg2002}
         {Cragg}, D.~M., {Sobolev}, A.~M., {Godfrey}, P.~D. 2002, \mnras, 331, 521

\bibitem[{{Cragg} {et~al.}(2005){Cragg}, {Sobolev}, & {Godfrey}}]{cragg2005}
         {Cragg}, D.~M., {Sobolev}, A.~M., {Godfrey}, P.~D. 2005, \mnras, 360, 533

\bibitem[{{Fujisawa} {et~al.}(2014){Fujisawa}, {Aoki}, {Nagadomi}, {Kimura}, {Shimomura}, {Takase}, {Sugiyama}, {Motogi}, {Niinuma}, {Hirota}, & {Yonekura}}]{fujisawa2014}
         {Fujisawa}, K., {Aoki}, N., {Nagadomi}, Y., {et~al.} 2014, \pasj, 66, 109

\bibitem[{{Fujisawa} {et~al.}(2015){Fujisawa}, {Yonekura}, {Sugiyama}, {Horiuchi}, {Hayashi}, {Hachisuka}, {Matsumoto}, & {Niinuma}}]{fujisawa2015}
         {Fujisawa}, K., {Yonekura}, Y., {Sugiyama}, K., {et~al.} 2015, The Astronomer's Telegram, 8286

\bibitem[{{Goedhart} {et~al.}(2004){Goedhart}, {Gaylard}, & {van der Walt}}]{goedhart2004}
         {Goedhart}, S., {Gaylard}, M.~J., {van der Walt}, D.~J. 2004, \mnras, 355, 553

\bibitem[{{Hu} {et~al.}(2016){Hu}, {Menten}, {Wu}, {Bartkiewicz}, {Rygl}, {Reid}, {Urquhart}, & {Zheng}}]{hu2016}
         {Hu}, B., {Menten}, K.~M., {Wu}, Y., {Bartkiewicz}, A., {Rygl}, K., {Reid}, M.~J., {Urquhart}, J.~S., {Zheng}, X. 2016, \apj, 833, 18

\bibitem[{{Hunter} {et~al.}(2017){Hunter}, {Brogan}, {MacLeod}, {Cyganowski}, {Chandler}, {Chibueze}, {Friesen}, {Indebetouw}, {Thesner}, & {Young}}]{hunter2017}
         {Hunter}, T.~R., {Brogan}, C.~L., {MacLeod}, G., {et~al.} 2017, \apj, 837, 29

\bibitem[{{Hunter} {et~al.}(2018){Hunter}, {Brogan}, {MacLeod}, {Cyganowski}, {Chibueze}, {Friesen}, {Hirota}, {Smits}, {Chandler}, & {Indebetouw}}]{hunter2018}
         {Hunter}, T.~R., {Brogan}, C.~L., {MacLeod}, G., {et~al.} 2018, \apj, 854, 170

\bibitem[{{Johnstone} {et~al.}(2013){Johnstone}, {Hendricks}, {Herczeg}, & {Bruderer}}]{johnstone2013}
         {Johnstone}, D., {Hendricks}, B., {Herczeg}, G.~J., {Bruderer}, S. 2013, \apj, 765, 133

\bibitem[{{Menten} (1991){Menten}}]{menten1991} {Menten}, K.~M. 1991, \apj, 380, L75

\bibitem[{{Meyer} {et~al.}(2017){Meyer}, {Vorobyov}, {Kuiper}, & {Kley}}]{meyer2017}
         {Meyer}, D.~M.-A., {Vorobyov}, E.~I., {Kuiper}, R., {Kley}, W. 2017, \mnras, 464, L90

\bibitem[{{Moscadelli} {et~al.}(2017){Moscadelli}, {Sanna}, {Goddi}, {Walmsley}, {Cesaroni}, {Caratti o Garatti}, {Stecklum}, {Menten}, & {Kraus}}]{moscadelli2017}
         {Moscadelli}, L., {Sanna}, A., {Goddi}, C., {et~al.} 2017, \aap, 600, L8

\bibitem[{{Pietka} {et~al.}(2017){Pietka}, {Staley}, {Pretorius}, & {Fender}}]{pietka2017}
         {Pietka}, M., {Staley}, T.~D., {Pretorius}, M.~L., {Fender}, R.~P. 2017, \mnras, 471, 3788

\bibitem[{{Sobolev} {et~al.}(1997){Sobolev}, {Cragg}, & {Godfrey}}]{sobolev1997}
         {Sobolev}, A.~M., {Cragg}, D.~M., {Godfrey}, P.~D. 1997, \aap, 324, 211

\bibitem[{{Stecklum} {et~al.}(2016){Stecklum}, {Caratti o Garatti}, {Cardenas}, {Greiner}, {Kruehler}, {Klose}, & {Eisloeffel}}]{stecklum2016}
         {Stecklum}, B., {Caratti o Garatti}, A., {Cardenas}, M.~C., {et~al.} 2016, The Astronomer's Telegram, 8732

\bibitem[{{Surcis} {et~al.}(2013){Surcis}, {Vlemmings}, {van Langevelde}, {Hutawarakorn Kramer}, & {Quiroga-Nunez}}]{surcis2013}
         {Surcis}, G., {Vlemmings}, W.~H.~T., {van Langevelde}, H.~J.; {Hutawarakorn Kramer}, B., {Quiroga-Nunez}, L.~H. 2013, \aap, 556, A73

\bibitem[{{Szymczak} {et~al.}(2000){Szymczak}, {Hrynek}, & {Kus}}]{szymczak2000}
         {Szymczak}, M., {Hrynek}, G., {Kus}, A.~J. 2000, \aaps, 143, 269

\bibitem[{{Szymczak} {et~al.}(2012){Szymczak}, {Wolak}, {Bartkiewicz}, & {Borkowski}}]{szymczak2012}
         {Szymczak}, M., {Wolak}, P., {Bartkiewicz}, A., {Borkowski}, K.~M. 2012, Astron.~Nachr., 333, 634

\bibitem[{{Szymczak} {et~al.}(2018){Szymczak}, {Olech}, {Sarniak}, {Wolak}, & {Bartkiewicz}}]{szymczak2018}
         {Szymczak}, M., {Olech}, M., {Sarniak}, R., {Wolak}, P., {Bartkiewicz}, A. 2018, \mnras, 474, 219

\bibitem[{{Wang} {et~al.}(2011){Wang}, {Beuther}, {Bik}, {Vasyunina}, {Jiang}, {Puga}, {Linz}, {Rod\'on}, {Henning}, & {Tamura}}]{wang2011}
          {Wang}, Y., {Beuther}, H., {Bik}, A., {et~al.} 2011, \aap, 527, A32

\bibitem[{{Zinchenko} {et~al.}(2012){Zinchenko}, {Liu}, {Su}, {Kurtz}, {Ojha}, {Samal}, & {Ghosh}}]{zinchenko2012}
         {Zinchenko}, I., {Liu}, S.-Y., {Su}, Y.-N., {et~al.} 2012, \apj, 755, 177

\bibitem[{{Zinchenko} {et~al.}(2015){Zinchenko}, {Liu}, {Su}, {Salii}, {Sobolev}, {Zemlyanukha}, {Beuther}, {Ojha}, {Samal}, & {Wang}}]{zinchenko2015}
         {Zinchenko}, I., {Liu}, S.-Y., {Su}, Y.-N., {et~al.} 2015, \apj, 810, 10

\bibitem[{{Zinchenko} {et~al.}(2017){Zinchenko}, {Liu}, {Su}, & {Zemlyanukha}}]{zinchenko2017}
         {Zinchenko}, I., {Liu}, S.-Y., {Su}, Y.-N., {Zemplyanukha}, P. 2017, arXiv:1711.05936



\end{thebibliography}

\begin{appendix}
\section{Description of the burst phase}
After MJD $\sim$56716, several new features appeared in the velocity range of 0.6 to 8.7\,\kms\, , while the preexisting features began to increase in
intensity (Fig.~\ref{sel-LC-int}).

New emission appeared at velocities lower than 1.60\,\kms\, and increased by a factor of 6-10 during $\sim$550-600~days, then rapidly decreased over 30-35~days
to less than 1.5-2\,Jy around MJD 57330, and during the next $\sim$100-120\,days, it was below our detection limit of $\sim$0.9\,Jy; finally, it reappeared
for a period of $\le$510\,days. The emission in the velocity ranges of 0.55-1.03 and 1.07-1.60\,\kms\, (Fig.~\ref{sel-LC-int}a,\,b) dropped below the detection limit after
MJD 57760 and 57940, respectively. A similar light curve applies for the emission from 1.65 to 2.04\,\kms\, (Fig.~\ref{sel-LC-int}c), with the exception that a weak feature
was present before MJD 56716 and the emission was still present at the end of the observations. The emission in these three velocity intervals showed evidence
of flickering-like events (e.g., \citealt{fujisawa2014}; \citealt{szymczak2018}) that could not be fully sampled with our observation cadence. As the emission in
other velocity intervals did not show similar behavior, we are confident that these effects are not spurious features owing to calibration errors.

In the velocity range of 2.08 to 2.70\,\kms\, (Fig.~\ref{sel-LC-int}d), the emission presented a rather flat course over the period $\sim$MJD 56716-57318. It then rapidly dropped
to a level of 2-4$\sigma$ above noise level around MJD 57344 and monotonically increased since MJD $\sim$57680.

The emission from 2.74 to 3.14\,\kms\, (Fig.~\ref{sel-LC-int}e) was constant before MJD $\sim$56716, then exponentially increased by a factor of four, reaching
a maximum around MJD 57270 and rapidly decreased below the pre-burst level around MJD 57360, then reappeared after $\sim$120\,days and reached
the second maximum around MJD 57760. Just before the end of the observations, it reached the pre-burst level. The emission from 3.18 to 3.80\,\kms\,
(Fig.~\ref{sel-LC-int}f) showed a similar behavior, with the exception that two spikes occurred at MJD 57279 and 57314 and the second very flat maximum began at MJD 57553
and lasted $\sim$137\,days.

The persistent emission from 3.84 to 4.28\,\kms\, (Fig.~\ref{sel-LC-int}g) peaked at 4.10\,\kms\, experienced a decay by a factor of five since MJD $\sim$56716 to 57349,
then it increased at its maximum at the pre-flare level around MJD 57727, and finally, it decreased again.

The strongest pre-flare persistent feature at 4.63\,\kms\, slightly increased since MJD $\sim$56716 and displayed a variability of 60\%.  Since MJD 57391,
the flux density integrated over the velocity range 4.32-4.72\,\kms\, (Fig.~\ref{sel-LC-int}h) decreased by a factor of 4.9 during 64\,days.

The emission from 4.76 to 5.16\,\kms\, (Fig.~\ref{sel-LC-int}i) slightly increased around MJD 56751, then remained at the pre-flare level until MJD 57263, when the integrated
flux density began to grow by a factor of 5.6 during 74\,days. In the high-intensity state, the emission showed moderate (30\%) variations over $\sim$5\,months, then
exponentially decreased, showing at the end of observations an intensity a factor of two lower than in the pre-flare state.

The emission of 5.20-5.38\,\kms\, (Fig.~\ref{sel-LC-int}j) diminished by a factor of two in the interval of MJD 56716-57260, then increased by a factor of 10 over $\sim$55\,days,
and its maximum occurred around MJD 57477. The emission exhibited variability of up to 90\% in the high state and moderate (30\%) variability during an exponential decline
phase; the intensity at the end of the observations was a factor of 3.1 lower than the pre-flare value.

The emission from 5.42 to 5.60\,\kms\, (Fig.~\ref{sel-LC-int}k) increased by a factor of 2.5 in the period of MJD 56716-57279, then declined to the pre-flare intensity
around MJD 57298, and after $\sim$40\,days, it reached a maximum intensity a factor of 4.6 greater than the pre-flare value. In the high state, the emission showed variability
of up to 70\%, and exponentially decreased, reaching a flat dip around MJD 57840, then slightly increased at the end of the monitoring.

The emission from 5.63 to 5.82\,\kms\, (Fig.~\ref{sel-LC-int}l) exponentially increased in the period of MJD 56706 to 57320 when the integrated flux density increased by
a factor of 37 with a dip around MJD 57303. Since MJD 57320, the intensity dropped by a factor of 4.5 during 56\,days, then showed moderate ($\sim$40\%) variation
with a flat maximum around MJD 57681. The intensity at the end of the observations was still a factor of 4.4 higher that the pre-flare value.
A similar light curve was seen for the emission in the velocity range from 5.86 to 6.08\,\kms\, (Fig.~\ref{sel-LC-int}m), with the exception
that the peak intensity was a factor of 240 higher than the pre-flare value.

New emission at 6.12-6.61\,\kms\, (Fig.~\ref{sel-LC-int}n) appeared around MJD 56794, reaching a maximum around MJD 57338; the integrated flux density increased by more
than a factor of 1130, then an exponential decrease occurred with a characteristic time of 420\,days. The emission at 6.65 to 7.05\,\kms\, (Fig.~\ref{sel-LC-int}o)
appeared around MJD 57023 and after $\sim$336\,days reached its maximum; the intensity increased by more than a factor of 480, then exponentially decreased with
a characteristic time of 145\,days. This emission at a level of 10$\sigma$ was still present at the end of the observations.
A similar burst profile was seen for the new emission in the velocity range of 7.09-7.26\,\kms\, (Fig.~\ref{sel-LC-int}p), which appeared at MJD $\sim$57254, and its
intensity at the maximum (MJD 57339) rose by more than a factor of 8. Moreover, in the high state, it varied with a relative amplitude of up to 2.1 on timescales of
10-30\,days. This emission dropped below the detection level around MJD 57980.

New emission in the velocity interval of 7.30 to 8.01\,\kms\, (Fig.~\ref{sel-LC-int}q) appeared around MJD 57131 and exponentially increased, reaching a maximum
around MJD 57319, then rapidly declined; the width of the spike profile at half intensity was only $\sim$21\,days. The second maximum occurred around MJD 57704 and then
the emission displayed a slow exponential decline and remained above the detection limit at the end of the monitoring.

In the velocity range of 8.06 to 8.67\,\kms\, (Fig.~\ref{sel-LC-int}\,r), new emission appeared at MJD 57280 and exponentially increased by a factor of 45 during 52\,days, then showed
fluctuations by a factor of 1.8-4.4 over $\sim$100\,days. It peaked at MJD 57411, and then exponentially disappeared after 375\,days.
\end{appendix}

\end{document}